\documentclass[a4paper,fleqn]{cas-sc}
\usepackage[numbers]{natbib}
\usepackage{booktabs}
\usepackage{adjustbox}
\usepackage{longtable}
\usepackage{caption}
\usepackage{mathtools}
\usepackage{amssymb}
\usepackage{amsmath}
\usepackage{etoolbox}
\usepackage{graphicx}
\usepackage[utf8]{inputenc}
\usepackage[T1]{fontenc}
\usepackage[english]{babel}

\makeatletter
\pretocmd{\start@align}{\setlength{\mathindent}{0pt}}{}{}
\makeatother

\newlength{\myalignwidth}

\def\tsc#1{\csdef{#1}{\textsc{\lowercase{#1}}\xspace}}
\tsc{WGM}
\tsc{QE}
\tsc{EP}
\tsc{PMS}
\tsc{BEC}
\tsc{DE}

\begin{document}
	\let\WriteBookmarks\relax
	\def\floatpagepagefraction{1}
	\def\textpagefraction{.001}
	\let\printorcid\relax
	
	\shorttitle{The survival of the weakest in a donation game}
	
	\shortauthors{C. Wang, J. Li, X. Wang {\it et~al.}}
	
	\title [mode = title]{The survival of the weakest in a biased donation game}                      
	
	\author[1]{Chaoqian Wang}
	\ead{CqWang814921147@outlook.com}
	\credit{xxx}
	
	\author[2]{Jingyang Li}
	\ead{fifthocean@sjtu.edu.cn}
	
	\author[3]{Xinwei Wang}
	\cormark[1]
	\cortext[cor1]{Corresponding author}
	\ead{wangxinwei@dlut.edu.cn}
	\credit{xxx}
	
	\author[4,5,6]{Wenqiang Zhu}
	\ead{wqzhu@buaa.edu.cn}
	\credit{xxx}
	
	\author[7]{Attila Szolnoki}
	\ead{szolnoki.attila@ek-cer.hu}
	\credit{xx}
	
	\address[1]{School of Mathematics and Statistics, Nanjing University of Science and Technology, Nanjing 210094, China}
	\address[2]{Zhiyuan College, Shanghai Jiao Tong University, Shanghai, China}
	\address[3]{Department of Engineering Mechanics, State Key Laboratory of Structural Analysis, Optimization and CAE Software for Industrial Equipment, Dalian University of Technology, Dalian, 116024, China}
	\address[4]{School of Artificial Intelligence, Beihang University, Beijing 100191, China}
	\address[5]{Key Laboratory of Mathematics, Informatics and Behavioral Semantics, Beihang University, Beijing 100191, China}
	\address[6]{Zhongguancun Laboratory, Beijing 100094, China}
	\address[7]{Institute of Technical Physics and Materials Science, Centre for Energy Research, P.O. Box 49, H-1525 Budapest, Hungary}

	\begin{abstract}
		Cooperating first then mimicking the partner's act has been proven to be effective in utilizing reciprocity in social dilemmas.
		However, the extent to which this, called Tit-for-Tat strategy, should be regarded as equivalent to unconditional cooperators remains controversial. Here, we introduce a biased Tit-for-Tat (T) strategy that cooperates differently toward unconditional cooperators (C) and fellow T players through independent bias parameters. The results show that, even under strong dilemmas in the donation game framework, this three-strategy system can exhibit diverse phase diagrams on the parameter plane. In particular, when T-bias is small and C-bias is large, a ``hidden T phase'' emerges, in which the weakest T strategy dominates. The dominance of the weakened T strategy originates from a counterintuitive mechanism characterizing non-transitive ecological systems: T suppresses its relative fitness to C, rapidly eliminates the cyclic dominance clusters, and subsequently expands slowly to take over the entire population. Analysis in well-mixed populations confirms that this phenomenon arises from structured populations. Our study thus reveals the subtle role of bias regulation in cooperative modes by emphasizing the ``survival of the weakest'' effect in a broader context.
	\end{abstract}
	
	
	

	\maketitle

	\section{Introduction}
	
	Those who are willing to act to benefit others at a cost to themselves are obviously in a disadvantageous position compared to those who refuse to do so~\cite{wang_x_epl20,sun_xp_pla25}. This is the essence of the social dilemma, in which cooperative and defective strategies compete in an evolutionary process~\cite{maynard_82}. An escape route from the trap of an undesirable outcome---where all actors choose the seemingly rational option of defection---may involve the application of different kinds of incentives~\cite{chen_xj_jrsi14,feng_m_csf23,li_k_csf21,liu_lj_srep17}. These incentives, including punishment for defection or rewarding of cooperation, can make the payoff of cooperators competitive with that of defectors~\cite{an_x_amc26}. However, such incentives are generally costly, transforming the original dilemma into a new stage~\cite{liu_lj_rsif22,jiang_ll_jpa25,liu_yy_jrsif25}.
	
	An alternative way to address this problem is to introduce a third strategy whose presence may modify the competitive ranking and reduce the apparent advantage of defectors. Counterintuitively, the presence of a neutral party, which does not actively participate in the original conflict, can help maintain a certain level of cooperation even under harsh conditions~\cite{hauert_s02}. These actors, often referred to as loners, neither cooperate nor allow themselves to be exploited by defectors, and instead earn a fixed but modest income. This makes them vulnerable to cooperators while simultaneously allowing them to outperform defectors. As a result, a non-transitive dominance relationship emerges that sustains the diversity of competing strategies~\cite{szolnoki_pre09b,tao_yw_epl21}.
	
	Interestingly, a similar non-transitive relationship arises in social dilemma situations when players can adopt a responsive strategy beyond unconditional cooperation or defection. This is the well-known Tit-for-Tat strategy, proposed by Anatol Rapoport, which cooperates with cooperators but defects against defectors~\cite{axelrod_84}. The presence of this more sophisticated strategy also results in oscillatory dynamics among three competing strategies~\cite{nowak_n92a}. At this point, it is worth emphasizing that this kind of cyclic dominance is a more general phenomenon and has been identified as a potential explanation for biodiversity in ecological systems~\cite{kerr_n02,szolnoki_epl20,szolnoki2014cyclic,szolnoki_njp15}: there is no single winner, but all participants in an invasion loop can persist~\cite{park_csf23}. Notably, non-transitive systems can also produce counterintuitive effects~\cite{szolnoki_njp14}. One such effect is the so-called ``survival of the weakest'' phenomenon~\cite{frean_prsb01}, first reported by Kei-ichi Tainaka~\cite{tainaka_pla95}. Specifically, when the fitness of one competitor in a cyclic interaction loop is reduced, the main loser of this intervention is often its original predator~\cite{avelino_pre19b,szolnoki_csf20b}.
	
	We now
	merge the aforementioned research directions and examine whether similar system behavior can be observed in a basic social dilemma setting. More precisely, we consider the traditional donation game, in which some players (C) are willing to pay a personal cost to ensure a benefit to their partner, while others refuse to do so and defect (D). Beyond this traditional two-strategy framework, we introduce a third strategy that, in contrast to cooperators, does not support defectors, but instead supports only C and T players. This strategy behaves conceptually similarly to the aforementioned Tit-for-Tat strategy and is therefore denoted as the T strategy.
	
	More importantly, the strategy we introduce represents a generalization of the traditional Tit-for-Tat strategy, as it allows for different levels of support toward C and T partners. In other words, we employ a biased Tit-for-Tat strategy, which enables us to explore a broader and more comprehensive parameter space that determines the relationships among competing strategies. Our aim is to investigate the resulting system dynamics and identify the potential consequences arising from the presence of this newly introduced third strategy.

	\section{Model}\label{sec_model}
	In the traditional donation game, players can basically cooperate (marked by C) or defect (denoted by D). A cooperator donates an amount $c$, providing the opponent with an amplified benefit $b$ ($b > c$), while a defector donates nothing. This setup constitutes a specific social dilemma situations, called prisoner's dilemma. Traditionally, the dilemma strength is related to the ratio $b/c$; however, the cost-to-benefit ratio $c/b$ is a preferable measure~\cite{fotouhi2018conjoining}, owing to its monotonicity and bounded domain. This alternative has gained traction in recent studies~\cite{hauert2025phase}. Following this, we define $r = c/b$. By a normalization of $1/b$, we equivalently describe the game such that a cooperator donates $r$ (where $0 < r < 1$), and the opponent receives a benefit of 1.
	
	The Tit-for-Tat (T) is the third strategy type, which donates only to cooperators. A question arises as to whether the T strategy itself qualifies as an equivalent cooperator. Here, we assume that the T strategy applies biased donations to C and fellow T strategies. The biases towards C and T are parameterized by $  \theta_\text{C}  $ and $  \theta_\text{T}  $ ($  \theta_\text{C}, \theta_\text{T} > 0  $), respectively, which serve as coefficients on the donation amount. That is, a T-player donates $  \theta_\text{C} c  $ to a C-player and $  \theta_\text{T} c  $ to a T-player. We maintain the fixed benefit-to-cost ratio $  b/c  $, which means that C and T strategies receive $  \theta_\text{C} b  $ and $  \theta_\text{T} b  $, respectively. Again, through normalization by $  1/b  $, the T strategy can be equivalently described as donating $  \theta_\text{C} r  $ to a C-player or $  \theta_\text{T} r  $ to a T-player, with the opponent receiving $  \theta_\text{C}  $ and $  \theta_\text{T}  $, respectively. Against defectors, T-players exhibit no bias and still donate nothing to them. 
	
\begin{figure}
		\centering
		\includegraphics[width=.55\textwidth]{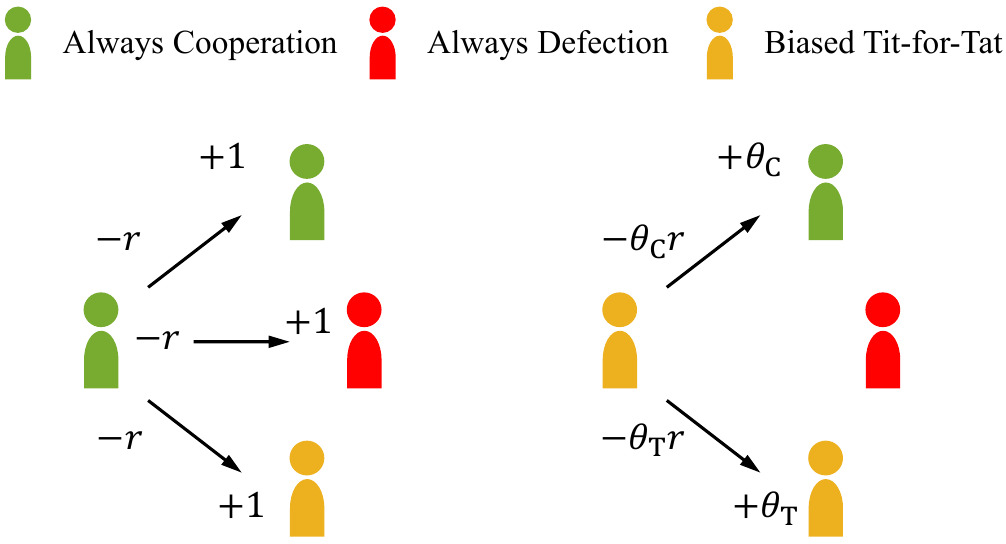}
		\caption{Schematic illustration of the three-strategy game system. Left: A C-player pays a cost $-r$ to provide $+1$ to the partner. Right: A T-player's behavior depends on the opponent's strategy: when facing a C-player, it pays $-\theta_\text{C} r$ to confer a benefit $+\theta_\text{C}$; when facing a T-player, it pays $-\theta_\text{T} r$ to confer a benefit $+\theta_\text{T}$.}
		\label{fig_demo}
	\end{figure}

	We denote the strategy indices of C, D, and T as 1, 2, and 3, respectively. The payoff that player $  i  $ gains when interacting with player $  j  $ can then be expressed as $  \mathbf{M}_{s_i s_j}  $, where $  s_i, s_j \in \{1, 2, 3\}  $ are the strategy indices of players $  i  $ and $  j  $, and $  \mathbf{M}  $ is the payoff matrix, 
	\begin{equation}\label{eq_M}
		\mathbf{M}=
		\begin{pmatrix}
			1-r & -r & \theta_\text{C} -r \\
			1 & 0 & 0 \\
			1-\theta_\text{C} r & 0 & \theta_\text{T} (1-r)
		\end{pmatrix}.
	\end{equation}
	A schematic illustration of this three-strategy system is shown in Fig.~\ref{fig_demo}.
	
	Consider a network structure of an $  L \times L  $ lattice with periodic boundary conditions, where every site represent an actor who interacts with its $ k=4  $ nearest neighbors. The latter is denoted by $  \Omega_i  $. Initially, the three available strategies are distributed randomly among the players. In a basic simulation cycle, 
	we calculate average payoff $\pi_i$ of a randomly selected actor $i$
	from interactions with its neighbors $l\in \Omega_i$, 
	\begin{equation}
		\pi_i=\frac{1}{k}\sum_{l\in \Omega_i} \mathbf{M}_{s_i s_l}.
	\end{equation}
	Then, a neighbor $j$ is chosen randomly, whose average payoff $  \pi_j  $ is computed in the same manner. The likelihood of player $i$ imitates player $j$ can be calculated as 
	\begin{equation}
		W_{s_i\gets s_j}=\frac{1}{1+\exp{(-(\pi_j-\pi_i)/K)}},
	\end{equation}
	where $K = 0.1$ represents a moderate noise level that ensures a certain degree of randomness when imitating higher-payoff strategies.
	
	\section{Results}
	
	\begin{figure}
		\centering
		\includegraphics[width=.8\textwidth]{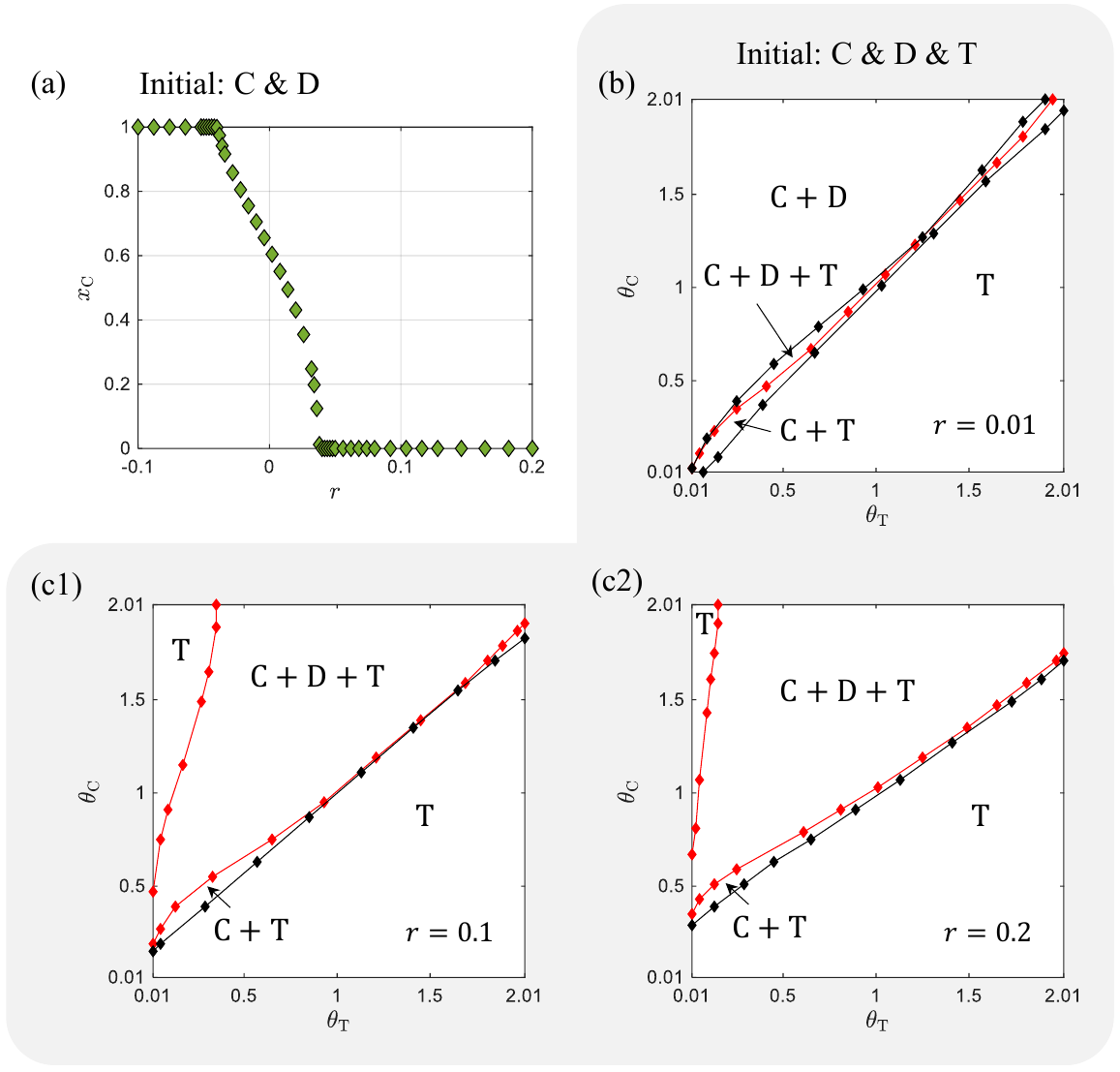}
		\caption{System behavior under different dilemmas. (a) In the traditional two-strategy donation game, increasing $r$ reduces the level of cooperation~\cite{hauert2025phase}. (b) When cooperation can emerge in the traditional setting ($r = 0.01$), the $\theta_\text{T}$-$\theta_\text{C}$ parameter plane exhibits a C+D phase, a coexistence C+D+T phase, a C+T phase, and a T phase. (c1), (c2) Interestingly, when cooperation in the C+D phase is suppressed in the traditional two-strategy games ($r = 0.1$ and $r = 0.2$), a hidden T phase emerges for small T-bias $\theta_\text{T}$ and large C-bias $\theta_\text{C}$. The red lines indicate the emergence or extinction of the D strategy.}
		\label{fig_phase}
	\end{figure}
	
\begin{figure}
		\centering
		\includegraphics[width=.95\textwidth]{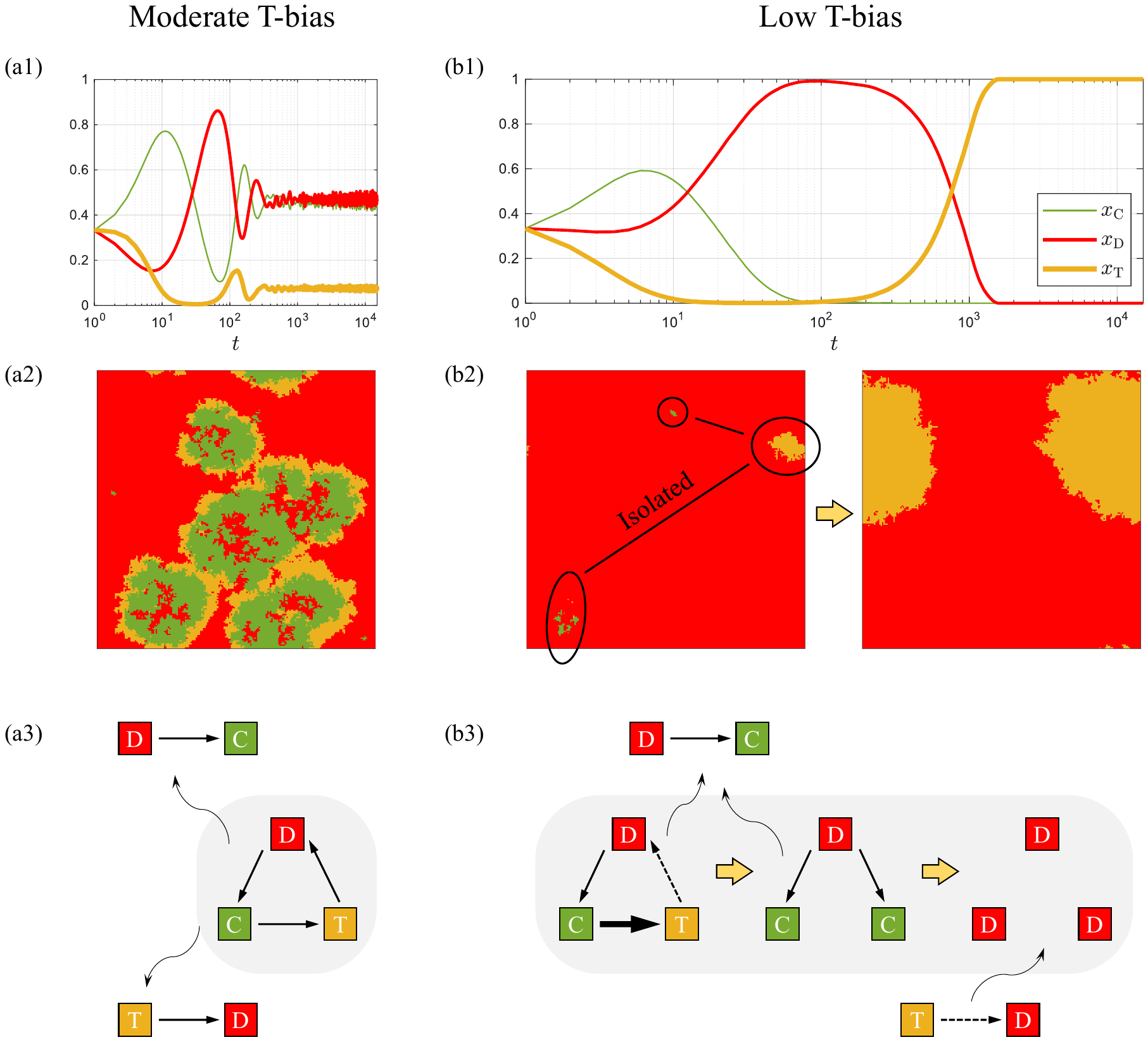}
		\caption{Spatial dynamics underlying the coexistence phase and the hidden T phase. (a1)--(a3) Under moderate T-bias ($\theta_\text{T} = 1$), the coexistence of all available strategies is sustained by their intransitive relation described as 
			$\text{D} \to \text{C} \to \text{T} \to \text{D}$. (b1)--(b3) Under low T-bias ($\theta_\text{T} = 0.1$), the hidden T phase emerges through a two-step mechanism: cyclic-dominance clusters first die out, after which isolated T individuals slowly expand. (a1), (b1) Typical time evolution of the strategy frequencies in a $600 \times 600$ population for the two phases. (a2), (b2) Snapshots of macroscopic spatial patterns in a $200 \times 200$ population for the two phases. (a3), (b3) Schematic illustrations of the microscopic spatial dynamics for the two phases. The arrows indicate the invasion direction between strategies. Other parameters: $r=0.1$, $\theta_\text{C}=1.5$.}
		\label{fig_snap}
	\end{figure}

	In the simulations, the three strategies are initially distributed randomly with equal probability. In a complete Monte Carlo circle we apply the previously defined basic round $L^2$ times,
	in which a randomly selected individual may update current strategy according to the procedure described in Section~\ref{sec_model}. We perform simulations for at least $1.5 \times 10^4$ MCS and record the fractions of the three possible strategies ($x_\text{C}$, $x_\text{D}$, and $x_\text{T}$). The stationary values of strategy frequencies are obtained by averaging over the final $5 \times 10^3$ MCS (near critical phase transitions, we extend the simulation to $3 \times 10^4$ MCS and average over the last $1.5 \times 10^4$ MCS). To minimize finite-size effects, all simulations (except for illustrative snapshots) are conducted on lattices of size $L = 600$; for visual demonstration of spatial patterns, we use $L = 200$ (Fig.~\ref{fig_snap}).
	
	Simulations initialized with only the C and D strategies reproduce the classic results of the two-strategy donation game model [Fig.~\ref{fig_phase}(a)]~\cite{hauert2025phase}. As the cost-to-benefit ratio $r$ increases, the portion of cooperator agents decreases. Building on the distinct cooperative levels observed in the classic two-strategy model, we examine the changes in system behavior upon introducing the T strategy. When cooperation can emerge in the classic model ($r = 0.01$), four distinct phases are observed in the $\theta_\text{T}$-$\theta_\text{C}$ parameter plane when initialized with all three strategies [Fig.~\ref{fig_phase}(b)]. For small T-bias and large C-bias, our model retains the mixed phase of cooperators and defectors.
	As T-bias increases and C-bias decreases, the T strategy gradually replaces both C and D, leading to transitions from a $\text{C} + \text{D} + \text{T}$ coexistence phase, to a $\text{C} + \text{T}$ phase, and finally to a T phase. Here, the dynamics of the $\text{C} + \text{T}$ phase are straightforward, analogous to the C + D phase in the classic two-strategy model. When $\theta_\text{C} \neq 1$, strategies C and T effectively act as two different cooperative strategies (i.e., quasi-defective cooperation and quasi-cooperative cooperation)~\cite{wang2021public}, where quasi-cooperation survives within quasi-defection through the spatial reciprocity mechanism.
	
	When cooperation is suppressed in the classic model ($r = 0.1$ and $r = 0.2$), the $\theta_\text{T}$-$\theta_\text{C}$ parameter plane still exhibits regimes where only C and T coexist, as well as regions supporting all three strategies to survive [Fig.~\ref{fig_phase}(c1) and (c2)]. Surprisingly, a hidden T phase emerges for small T-bias and large C-bias, which is counterintuitive. Understanding the mechanism behind this hidden T phase builds upon insights into the coexistence mechanism of the $\text{C} + \text{D} + \text{T}$ phase. In the following, we discuss them separately.
	
	The mechanism underlying the $\text{C} + \text{D} + \text{T}$ coexistence phase is essentially rock-paper-scissors cyclic dominance [Fig.~\ref{fig_snap}(a1)--(a3)], which works under moderate T-bias. The cyclic dominance $\text{D} \to \text{C} \to \text{T} \to \text{D}$ is evident from the time evolution of the three strategy frequencies [Fig.~\ref{fig_snap}(a1)], where arrows indicate the direction of invasion. D invades C because, at high cost-to-benefit ratios $r$ considered here, cooperation cannot emerge in the traditional donation game. C invades T because, when C-bias is relatively larger than T-bias, the C strategy naturally gains an advantage over T. T invades D because individuals adopting D receive zero payoff against T, whereas individuals within T clusters obtain positive payoffs. This forms a closed loop that maintains dynamic stability when the mutual invasion rates remain within a balanced range. The typical spatial pattern of this cyclic dominance is shown in Fig.~\ref{fig_snap}(a2). Within regions dominated by D, isolated C or T individuals occasionally persist, but they rapidly go extinct on their own or become absorbed upon encountering propagating $\text{D} \to \text{C} \to \text{T} \to \text{D}$ cyclic waves, as illustrated in Fig.~\ref{fig_snap}(a2) and (a3).
	
	Building on this understanding, the hidden T phase under small T-bias and large C-bias can be explained [Fig.~\ref{fig_snap}(b1)--(b3)]. In brief, the T strategy reduces its own fitness relative to C, thereby eliminating the regions sustaining $\text{D} \to \text{C} \to \text{T} \to \text{D}$ cyclic dominance. The remaining T-clusters then slowly expand without interference from C and eventually take over all D strategies [Fig.~\ref{fig_snap}(b1)]. Within the parameter range supporting this hidden T phase, (1) C holds a strong advantage over T, and (2) T holds only a weak advantage over D. Here, we highlight two key aspects of the dynamics. First, the original balance of $\text{D} \to \text{C} \to \text{T} \to \text{D}$ cyclic dominance is disrupted by the combined effects of factors (1) and (2): an overly weak T is rapidly invaded and extinguished by C, breaking the cyclic loop and leaving C vulnerable to invasion by D [Fig.~\ref{fig_snap}(b3)]. Second, the rate of the isolated $\text{T} \to \text{D}$ invasion process is greatly slowed by factor (2), which decelerates the expansion of T clusters and, importantly, reduces the probability of encountering persisting $\text{D} \to \text{C} \to \text{T} \to \text{D}$ or $\text{D} \to \text{C}$ regions [Fig.~\ref{fig_snap}(b2), left]. Once these processes have died out, the remaining T clusters that avoided encounter continue their slow takeover of all D strategies [Fig.~\ref{fig_snap}(b2), right]. The hidden T phase thus emerges from the interplay of these two aspects.
	
	\begin{figure}
		\centering
		\includegraphics[width=.85\textwidth]{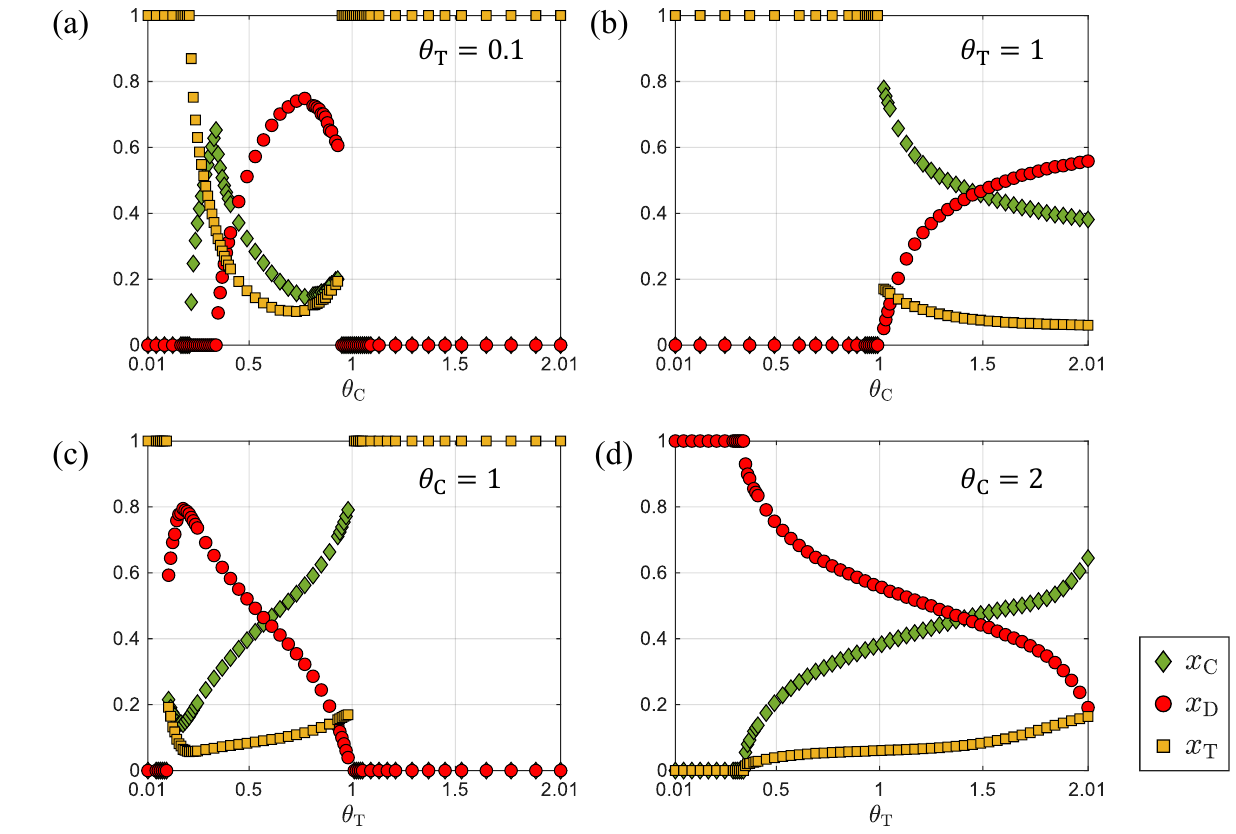}
		\caption{Extensive results for the stationary strategy frequencies as functions of the bias parameters. (a) For small T-bias ($\theta_\text{T} = 0.1$), moderate C-bias $\theta_\text{C}$ is most detrimental to cooperation. (b) For large T-bias ($\theta_\text{T} = 1$), increasing C-bias $\theta_\text{C}$ reduces cooperation. (c) For small C-bias ($\theta_\text{C} = 1$), moderate T-bias $\theta_\text{T}$ is most detrimental to cooperation. (d) For large C-bias ($\theta_\text{C} = 2$), increasing T-bias $\theta_\text{T}$ promotes cooperation. Other parameter: $r = 0.1$.}
		\label{fig_1D}
	\end{figure}
	
	Building upon the thorough understanding of the behavior in each phase, the diverse outcomes of the three-strategy system as well as the complex effects of the biased Tit-for-Tat strategy become interpretable. Fig.~\ref{fig_1D} presents extensive results illustrating the influence of C- and T-bias on cooperation in the system. When T-bias is small, both large and small C-bias promote cooperation, whereas moderate C-bias suppresses cooperation [Fig.~\ref{fig_1D}(a)]. When T-bias is large, increasing C-bias reduces cooperation [Fig.~\ref{fig_1D}(b)]. When C-bias is small, both large and small T-bias favor cooperation, whereas moderate T-bias hinders cooperation [Fig.~\ref{fig_1D}(c)]. When C-bias is large, increasing T-bias promotes cooperation [Fig.~\ref{fig_1D}(d)]. The mechanisms underlying these results have been elucidated in the preceding discussion of the phase diagrams.
	
	\begin{figure}
		\centering
		\includegraphics[width=.6\textwidth]{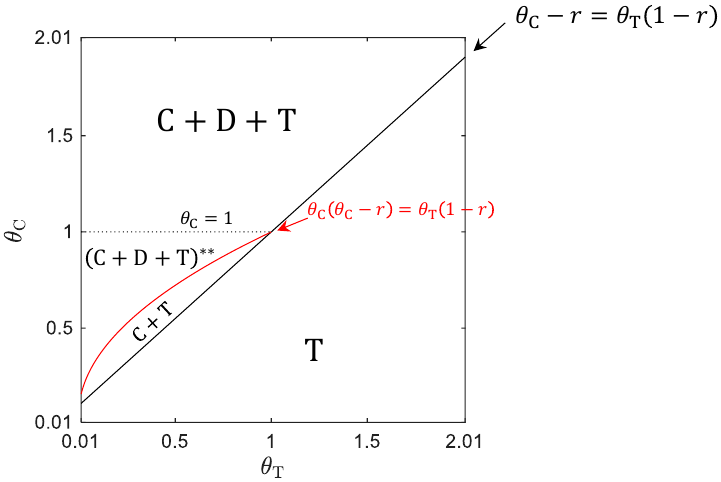}
		\caption{The hidden T phase does not exist in well-mixed populations. In the $\theta_\text{T}$-$\theta_\text{C}$ parameter plane, there are four theoretical phases in well-mixed populations: T (the T strategy dominates), $\text{C}+\text{T}$ (the C and T strategies coexist), $(\text{C}+\text{D}+\text{T})^{**}$ (the three strategies coexist and are stable at the interior equilibrium), and $\text{C}+\text{D}+\text{T}$ (the three strategies coexist in cyclic dominance). The T equilibrium is stable when $\theta_\text{T} (1 - r) > \max\{\theta_\text{C} - r, 0\}$; the $\text{C}+\text{T}$ equilibrium is stable when $\theta_\text{C}(\theta_\text{C} - r)<\theta_\text{T} (1 - r)<\theta_\text{C} - r$; the $(\text{C}+\text{D}+\text{T})^{**}$ equilibrium is stable when $\theta_\text{C}(\theta_\text{C} - r)>\theta_\text{T} (1 - r)$ and $\theta_\text{C}<1$; otherwise, all equilibria are unstable and the phase is $\text{C}+\text{D}+\text{T}$ in cyclic dominance. Numerical parameter: $r = 0.1$.}
		\label{fig_WM}
	\end{figure}
	
	Finally, we confirm that the hidden T phase observed at small T-bias is unique to structured populations and does not exist in well-mixed populations (Fig.~\ref{fig_WM}). As explained earlier, the dynamical mechanism of this hidden phase relies on spatial structure: T clusters suppress their own expansion rate, thereby avoiding encounters with cyclic-dominance clusters before the latter die out. In the dilemma regimes studied here, however, well-mixed populations reveal additional mathematical insights between the two coexistence phases of C, D, and T: when $\theta_\text{C}<1$, the three strategies coexist by stablizing at the interior equilibrium; when $\theta_\text{C}>1$, the three strategies coexist by cyclic dominance since all equilibria are unstable. Details of the theoretical analysis for this three-strategy system in well-mixed populations are provided in Appendix~\ref{sec_appen}.
	
	\section{Conclusion}
	The traditional Tit-for-Tat strategy distinguishes between cooperators and defectors by mirroring the opponent's behavior from the previous round, which exhibits robustness in repeated games~\cite{axelrod1981evolution}. However, when a T-player encounters another T-player, the level of cooperation may differ from the way it treats general cooperators: it can be either more generous or more conservative. Building on the normalized donation game, we extend this idea by introducing a biased Tit-for-Tat strategy, which allows independent bias coefficients $  \theta_\text{C}  $ and $  \theta_\text{T}  $ to be applied separately to the two types of cooperative partners (C and T), enabling a comprehensive exploration of the full parameter space.
	
	We find that even at harsh dilemma conditions, where cooperation is completely suppressed in the traditional donation game, appropriate tuning of the bias parameters can generate diverse phase diagrams. Beside the traditionally observed two-strategy phases of $\text{C} + \text{D}$ or $\text{C} + \text{T}$, and the state where all three competitors are present, the most striking phenomenon is the emergence of a hidden T phase in the region of small T-bias and large C-bias. In this region, the T strategy is the weakest among the three competitors, yet it ultimately survives and dominates the entire population.
	
	The essence of this hidden T phase is that the T strategy reduces its own fitness relative to C, driving its frequency down to an extremely low level. This process accelerates the extinction of the regions sustaining the $\text{D} \to \text{C} \to \text{T} \to \text{D}$ cyclic dominance. Subsequently, with no interference from C, the remaining isolated T clusters can expand slowly and eventually take over the entire population. This seemingly counterintuitive phenomenon is recognized as the so-called ``survival of the weakest'' effect that was previously observed in ecological cyclic dominance systems~\cite{tainaka1995indirect}. Our present work highlights that there is no need to presuppose non-transitive interactions, as it happens in Lotka--Volterra-type ecological systems, yet conceptually similar behavior emerges in social systems, where competitor interactions are more subtle. In particular, even the basic donation game can provide a framework in which we can observe this phenomenon. 
	
	It is important to stress that although this T phase represents a form of cooperation, it does not necessarily imply a high social welfare for the whole population. When $  \theta_\text{T}  $ is excessively small, the extremely low level of donation among fellow T individuals results in a substantial reduction in the total payoff of the population. This echoes a well-established insight in the literature: cooperation does not always equate to social welfare~\cite{hauert_s02,szolnoki_epl10,han2025cooperation,zhu2025evolution}. Our findings therefore highlight the subtle role that bias tuning plays in balancing cooperation against collective benefits in spatial games.
	
	Theoretical analysis of well-mixed populations further confirms that the hidden (weakest) T phase is unique to spatial structure. In a well-mixed population, the system supports only the (strongest) T phase, among others. The microscopic mechanism, that T clusters suppress their own expansion rate to avoid premature encounters with persisting cyclic-dominance regions, relies on spatial correlations and cannot emerge in a well-mixed environment where interactions are global. This is another nice example that illustrates the potential difference between well-mixed and structured populations~\cite{perc_pr17,sui2015evolutionary,li2021pool,wang_cq_nc24,zhu2024evolutionary,wang2026public}.
	
	For future research, the analysis of spatial dynamics under arbitrary selection strength remains challenging~\cite{ibsen2015computational}. From the perspective of invasion rates, the difficulty was reported in~\cite{kelsic2015counteraction,szolnoki2015vortices}. Under non-marginal selection, a faster invasion rate could be useful to beat a rival as reported in~\cite{perc2007cyclical}, but this is not always the case~\cite{szolnoki2024faster}. Instead, analytical investigations under marginal selection are promising and worthwhile directions~\cite{ohtsuki2006replicator}, which is not the case of our work. The present work offers a perspective for understanding the complexity of conditional cooperative strategies in spatial evolutionary games under non-marginal selection. 
	
	\section*{Acknowledgments}
	This work was supported by the National Research, Development and Innovation Office (NKFIH), Hungary under Grant No. K142948.
	
	\appendix
	\renewcommand\thefigure{\Alph{section}\arabic{figure}} 
	\renewcommand{\theequation}{A.\arabic{equation}}
	\setcounter{equation}{0}
	\section{Well-mixed populations}\label{sec_appen}
	\setcounter{figure}{0}
	We theoretically consider a well-mixed variation of 
	the proposed three-strategy model in an infinitely large population. 
	The fractions of the C, D, and T strategies are still denoted by $x_\text{C}$, $x_\text{D}$, and $x_\text{T}$, respectively, satisfying $x_\text{C} + x_\text{D} + x_\text{T} = 1$. Based on Eq.~(\ref{eq_M}), the average payoff values are given by
	\begin{equation}
		\begin{cases} 
			\displaystyle{\pi_\text{C}=x_\text{C}+\theta_\text{C} x_\text{T}-r}, \\
			\displaystyle{\pi_\text{D}=x_\text{C}}, \\
			\displaystyle{\pi_\text{T}=x_\text{C} (1-\theta_\text{C} r)+x_\text{T} \theta_\text{T} (1-r)}.
		\end{cases}
	\end{equation}
	The population-average payoff is
	\begin{equation}
		\langle\pi\rangle=x_\text{C} (x_\text{C}+\theta_\text{C} x_\text{T}-r)+(1-x_\text{C}-x_\text{T} ) x_\text{C}+x_\text{T} (x_\text{C} (1-\theta_\text{C} r)+x_\text{T} \theta_\text{T} (1-r)).
	\end{equation}
	
	According to the replicator dynamics~\cite{taylor1978evolutionary}, the time evolution of each strategy's frequency in the system is given by $\dot{x}_i = x_i (\pi_i - \langle \pi \rangle)$, where $i \in \{\text{C}, \text{D}, \text{T}\}$. This yields
	\begin{equation}\label{eq_dxdt}
		\begin{cases} 
			\displaystyle{\dot{x}_\text{C}=x_\text{C} (rx_\text{C}-\theta_\text{C} (1-r) x_\text{C} x_\text{T}+\theta_\text{C} x_\text{T}-\theta_\text{T} (1-r) x_\text{T}^2-r)}, \\
			\displaystyle{\dot{x}_\text{T}=x_\text{T} ((1-\theta_\text{C} )r x_\text{C}-\theta_\text{C} (1-r) x_\text{C} x_\text{T}+\theta_\text{T} (1-r) x_\text{T}-\theta_\text{T} (1-r) x_\text{T}^2 )}.
		\end{cases}
	\end{equation}
	Here, $x_\text{D} = 1 - x_\text{C} - x_\text{T}$ reduces the degrees of freedom of the system by one.
	
	To find the equilibria of system~(\ref{eq_dxdt}), it suffices to solve the equations $\dot{x}_\text{C} = 0$ and $\dot{x}_\text{T} = 0$. The stability of these equilibria can be analyzed by computing the eigenvalues $\lambda$ of the Jacobian $\mathbf{J}$ at each equilibrium point. The $\mathbf{J}$ matrix is given by
	\begin{equation}
		\mathbf{J}=\begin{pmatrix}
			2rx_\text{C}-2\theta_\text{C} (1-r) x_\text{C} x_\text{T}+\theta_\text{C} x_\text{T}-\theta_\text{T} (1-r) x_\text{T}^2-r
			& x_\text{C} (-\theta_\text{C} (1-r) x_\text{C}+\theta_\text{C}-2\theta_\text{T} (1-r) x_\text{T} ) \\
			x_\text{T} ((1-\theta_\text{C} )r-\theta_\text{C} (1-r) x_\text{T} )
			& (1-\theta_\text{C} )r x_\text{C}-2\theta_\text{C} (1-r) x_\text{C} x_\text{T}+2\theta_\text{T} (1-r) x_\text{T}-3\theta_\text{T} (1-r) x_\text{T}^2
		\end{pmatrix}.
	\end{equation}
	
	We find that system~(\ref{eq_dxdt}) has three vertex equilibria and at most one edge equilibrium and one interior equilibrium.
	
	The first equilibrium is $(x_\text{C}, x_\text{D}, x_\text{T}) = (1, 0, 0)$, for which the Jacobian matrix is
	\begin{equation}
		\mathbf{J}=\begin{pmatrix}
			r
			& \theta_\text{C} r \\
			0
			& (1-\theta_\text{C})r
		\end{pmatrix}.
	\end{equation}
	The eigenvalues of $\mathbf{J}$ are $\lambda_1 = r$ and $\lambda_2 = (1 - \theta_\text{C})r$. Since $\lambda_1 = r > 0$, the equilibrium $(1, 0, 0)$ is unstable.
	
	The second equilibrium is $(x_\text{C},x_\text{D},x_\text{T})=(0,1,0)$, for which the Jacobian matrix is
	\begin{equation}
		\mathbf{J}=\begin{pmatrix}
			-r
			& 0 \\
			0
			& 0
		\end{pmatrix}.
	\end{equation}
	The eigenvalues of $\mathbf{J}$ are $\lambda_1=-r$, $\lambda_2=0$. Since $\lambda_2 = 0$, the equilibrium $(0,1,0)$ is unstable.
	
	The third equilibrium is $(x_\text{C},x_\text{D},x_\text{T})=(0,0,1)$, for which the Jacobian matrix is
	\begin{equation}
		\mathbf{J}=\begin{pmatrix}
			\theta_\text{C}-r-\theta_\text{T} (1-r)
			& 0 \\
			-(\theta_\text{C}- r)
			& -\theta_\text{T} (1-r)
		\end{pmatrix}.
	\end{equation}
	The eigenvalues of $\mathbf{J}$ are $\lambda_1=\theta_\text{C}-r-\theta_\text{T} (1-r)$, $\lambda_2=-\theta_\text{T} (1-r)$. The equilibrium $(0, 0, 1)$ is stable whenever $\theta_\text{T} (1-r)>\max\{\theta_\text{C}-r,0\}$.
	
	The possible edge equilibrium $(x_\text{C}^*,0,x_\text{T}^*)$ (i.e., the $\text{C}+\text{T}$ phase) satisfies 
	\begin{equation}
		(x_\text{C}^*,0,x_\text{T}^*)
		=\left(
		\frac{\theta_\text{C}-r-\theta_\text{T} (1-r)}{(\theta_\text{C}-\theta_\text{T} )(1-r)} ,0,\frac{(1-\theta_\text{C} )r}{(\theta_\text{C}-\theta_\text{T} )(1-r)} 
		\right),
	\end{equation}
	for which the Jacobian matrix is 
	\begin{equation}
		\mathbf{J}=\begin{pmatrix}
			\dfrac{({\theta_\text{C}}^2- T)(\theta_\text{C}-r-\theta_\text{T} (1-r))r}{(\theta_\text{C}-\theta_\text{T} )^2 (1-r) }
			& \dfrac{(\theta_\text{C}  - 2\theta_\text{T}  +\theta_\text{C} \theta_\text{T} )(\theta_\text{C}-r-\theta_\text{T} (1-r))r}{(\theta_\text{C}-\theta_\text{T} )^2 (1-r) } \\[1em]
			-\dfrac{\theta_\text{T} (1-\theta_\text{C} )^2 r^2}{(\theta_\text{C}-\theta_\text{T} )^2 (1-r) }
			& \dfrac{(1-\theta_\text{C} )(\theta_\text{C} r - 2\theta_\text{T} r + {\theta_\text{T}}^2 r - (\theta_\text{C}-\theta_\text{T} )^2 )r}{(\theta_\text{C}-\theta_\text{T} )^2 (1-r) }
		\end{pmatrix}.
	\end{equation}
	The eigenvalues of $\mathbf{J}$ are 
	\begin{equation}
		\begin{cases}
			\lambda_1=\dfrac{(\theta_\text{C} (\theta_\text{C}- r)- \theta_\text{T} (1-r))r}{(\theta_\text{C}  - \theta_\text{T} )(1-r)},
			\\[1em]
			\lambda_2=-\dfrac{(1-\theta_\text{C} )(\theta_\text{C}  -\theta_\text{T}  -\theta_\text{T} (1-r))r}{(\theta_\text{C}  -\theta_\text{T} )(1-r)} =-(\theta_\text{C}-\theta_\text{T} )(1-r) x_\text{C}^* x_\text{T}^*.
		\end{cases}
	\end{equation}
	Since $0<x_\text{C}^*,x_\text{T}^*<1$ and $0<r<1$, the condition $\lambda_2<0$ is equivalent to $\theta_\text{C}>\theta_\text{T}$. On this basis, the additional condition for $\lambda_1<0$ to hold is $\theta_\text{C} (\theta_\text{C}- r)> \theta_\text{T} (1-r)$. Note that one prerequisite for the existence of 
	the solution
	$(x_\text{C}^*,0,x_\text{T}^*)$ is $\theta_\text{C}-r>\theta_\text{T} (1-r)$ (i.e., $x_\text{C}^*>0$), which contradicts the stability condition of $(0,0,1)$. Therefore, no bistability exists between them. It is readily seen that when $\theta_\text{C}>1$, the conditions $\lambda_1<0$ and $\lambda_2<0$ cannot be satisfied simultaneously. When $\theta_\text{C}<1$, the condition $\theta_\text{C}-r>\theta_\text{T} (1-r)$ (required for the existence of the boundary equilibrium) is stricter than $\theta_\text{C}>\theta_\text{T}$. In summary, the equilibrium $(x_\text{C}^*,0,x_\text{T}^*)$ exists and is stable iff $\theta_\text{C}(\theta_\text{C}-r)<\theta_\text{T} (1-r)<\theta_\text{C}-r$.
	
	The possible 
	solution
	of $(x_\text{C}^{**},x_\text{D}^{**},x_\text{T}^{**})$ (i.e., the $(\text{C}+\text{D}+\text{T})^{**}$ phase) satisfies 
	\begin{equation}
		(x_\text{C}^{**},x_\text{D}^{**},x_\text{T}^{**})
		=\left(
		\frac{\theta_\text{T} (1-r)}{\theta_\text{C}^2},\frac{\theta_\text{C}^2-\theta_\text{C} r-\theta_\text{T} (1-r)}{\theta_\text{C}^2},\frac{r}{\theta_\text{C}}  
		\right),
	\end{equation}
	for which the Jacobian matrix is 
	\begin{equation}
		\mathbf{J}=\begin{pmatrix}
			\dfrac{\theta_\text{T} (1-r) r^2}{{\theta_\text{C}}^2} 
			& \dfrac{\theta_\text{T} (1-r)({\theta_\text{C}}^2- \theta_\text{T}+ \theta_\text{T} r^2 )}{{\theta_\text{C}}^3}  \\[1em]
			-\dfrac{(\theta_\text{C}  - r) r^2}{\theta_\text{C}} 
			& \dfrac{\theta_\text{T} (1-r)(1-\theta_\text{C}+r)r}{{\theta_\text{C}}^2} 
		\end{pmatrix}.
	\end{equation}
	The eigenvalues of $\mathbf{J}$ are 
	\begin{equation}
		\begin{cases}
			\lambda_1=\dfrac{(-\theta_\text{T} (1-\theta_\text{C} )(1-r)+ \sqrt{D} )r}{2\theta_\text{C}^2},
			\\[1em]
			\lambda_2=\dfrac{(-\theta_\text{T} (1-\theta_\text{C} )(1-r)- \sqrt{D} )r}{2\theta_\text{C}^2},
		\end{cases}
	\end{equation}
	where
	\begin{equation}
		D=
		\theta_\text{T} (1-r)(\theta_\text{T}  - \theta_\text{T} r +\theta_\text{T}  {\theta_\text{C}}^2  + 4{\theta_\text{C}}^2 r - 4{\theta_\text{C}}^3+ 2\theta_\text{T} \theta_\text{C}  - 2\theta_\text{T} \theta_\text{C} r -\theta_\text{T}  {\theta_\text{C}}^2 r).
	\end{equation}
	It is readily seen that $\lambda_1>\lambda_2$. Therefore, when $D>0$, the equilibrium is stable if $\lambda_1<0$, i.e., $\theta_\text{C}(\theta_\text{C}-r)>\theta_\text{T} (1-r)$. When $D<0$, which corresponds to $\theta_\text{T}(1-r)<\left(\frac{2\theta_\text{C}}{1+\theta_\text{C}}\right)^2  (\theta_\text{C}-r)$, the stability condition requires the real parts of the eigenvalues to be negative, i.e., $\mathrm{Re}[\lambda_1]=\mathrm{Re}[\lambda_2]<0$, which implies $\theta_\text{C}<1$. When $\theta_\text{C}>1$, $\lambda_1$ is real and clearly $\lambda_1>0$; moreover, $\theta_\text{C}>1$ violates the stability condition $\theta_\text{C}<1$ for the case of complex eigenvalues. When $\theta_\text{C}<1$, note that $\left(\frac{2\theta_\text{C}}{1+\theta_\text{C}}\right)^2\leq \theta_\text{C}$, so the stability region is still given by $\theta_\text{C}(\theta_\text{C}-r)>\theta_\text{T} (1-r)$. In summary, the equilibrium $(x_\text{C}^{**},x_\text{D}^{**},x_\text{T}^{**})$ is stable iff $\theta_\text{C}(\theta_\text{C}-r)>\theta_\text{T} (1-r)$ and $\theta_\text{C}<1$.
	
	By summarizing the equilibria and their stability conditions above, we can obtain phase diagrams such as the one in Fig.~\ref{fig_WM}.
	

\end{document}